\documentclass[12pt]{article}

\usepackage{graphicx}
\usepackage{natbib}

\usepackage{journals}

\begin{document}

\title{Connecting VLBI and Gaia celestial reference frames}
\author{Zinovy Malkin$^{1,2,3}$\\
$^1$Pulkovo Observatory, St. Petersburg 196140, Russia \\
$^2$St.Petersburg State University, St. Petersburg, 199034 Russia \\
$^3$Kazan Federal University, Kazan 420000, Russia \\
e-mail: malkin@gao.spb.ru}
\date{\today}
\maketitle

\begin{abstract}
The current state of the link problem between radio and optical celestial reference frames is considered.
The main objectives of the investigations in this direction during the next few years are the preparation of a comparison
and the mutual orientation and rotation between the optical {\it Gaia} Celestial Reference Frame (GCRF)
and the 3rd generation radio International Celestial Reference Frame (ICRF3), obtained from VLBI observations.
Both systems, ideally, should be a realization of the ICRS (International Celestial Reference System) at micro-arcsecond level accuracy.
Therefore, the link accuracy between the ICRF and GCRF should be obtained with similar error level, which is not a trivial task
due to relatively large systematic and random errors in source positions at different frequency bands.
In this paper, a brief overview of recent work on the GCRF--ICRF link is presented.
Additional possibilities to improve the GCRF--ICRF link accuracy are discussed.
The suggestion is made to use astrometric radio sources with optical magnitude to 20$^m$ rather than to 18$^m$ as currently planned
for the GCRF--ICRF link.
In addition, the use of radio stars is also a prospective method to obtain independent and accurate orientation between the Gaia frame and the ICRF.
\end{abstract}


\section{Introduction}

The ESA's {\it Gaia} space astrometry mission \citep{Perryman2001, Lindegren2008} commenced successfully in December 2013 and its main
scientific program in July 2014.
One of the most important results of the {\it Gaia} mission will be a new highly-accurate optical celestial reference frame;
{\it Gaia} Celestial Reference Frame (GCRF).
Although the final GCRF version is expected to be available in the early 2020s, intermediate releases are planned, the first of them (DR1)
is expected to be released in 2016.
A new release of the VLBI-based celestial reference frame of similar accuracy, the 3rd realization of the International Celestial Reference Frame (ICRF3)
is planned for 2018 \citep{Jacobs2014}.

Both radio (ICRF) and optical (GCRF) frames must be realizations of the same concept of the International Celestial Reference System (ICRS), \citet{Arias1995}) with an expected accuracy at the level of a few tens of $\mu$as.
The link between the ICRF and GCRF should be realized at a similar level of accuracy, which is not a trivial task.
This problem is similar to that of the link between the {\it Hipparcos} Celestial Reference Frame (HCRF) and the ICRF \citep{Kovalevsky1997}.
Generally speaking, both GCRF and ICRF object positions are time-dependent.
Therefore, analogously to HCRF, both the orientation and rotation of the GCRF with respect to ICRF are to be defined. 
In this paper, the general term 'orientation' is used to avoid a non-principal discussion related to spin.
Interested readers can find more theoretical and practical details in \citet{Lindegren1995,Kovalevsky1997}.

On the one hand, the link task is more straightforward for the GCRF than for the HCRF, as most of the ICRF objects will be directly observed by {\it Gaia}.
On the other hand, the task is much more complicated due to the requirement that a much higher level of accuracy for the GCRF--ICRF link is needed so as to not compromise the high level of precision of the two frames. 

The basic method to tie the {\it Gaia} catalog to the ICRF, and hence to the ICRS, is using {\it Gaia} observations of compact extragalactic ICRF objects that have accurate radio astrometric positions. With the help of these common objects, the orientation angles between the ICRF and GCRF will be determined. Finally, the GCRF catalog will be aligned to the ICRS by applying these orientation angles. The accuracy of this link depends on many factors, such as random and systematic errors of both radio and optical catalogs.

The objective of this paper is to briefly overview recent work on the ICRF--GCRF link and to discuss new possibilities to improve the link accuracy. It should be noted that the link between the GCRF and ICRF is not a task currently planned for completion before the end of this decade \citep{Jacobs2014}.
Based on the {\it Hipparcos} experience, it can be envisioned that the work on improving such a link will be continued for a prolonged period after completion of the {\it Gaia} mission.
The improvements will be primarily based on a new VLBI-based celestial reference frame (CRF) realization of which the accuracy can improve over time.
New after-mission {\it Gaia} data reductions are also possible. Therefore, research and development of the methods for the linking of radio and optical reference frames will remain one of the primary tasks of fundamental astrometry throughout the next decade. 

The paper is structured as follows.
In section~\ref{sect:basic}, basic equations used to link two CRF realizations are described.
Section~\ref{sect:overview} contains a brief overview of recent investigations regarding aspects concerning the ICRF--GCRF link. 
The following three sections are devoted to a discussion of new possibilities and possible improvements in both theoretical analysis and the
final ICRF--GCRF link accuracy, such as the choice of the ICRF realization used for modelling and simulation (Section~\ref{sect:radio_frame}),
using more link sources (Section~\ref{sect:more_sources}), using radio stars (Section~\ref{sect:radio_stars}),
and proper accounting for the galactic aberration in proper motions (Section~\ref{sect:ga}).


\section{Basic equations}
\label{sect:basic}
 
Each catalog of the positions of celestial objects (CRF realization), be it ICRF or GCRF, represents its own coordinate frame linked to the ICRS at some degree of accuracy.
Mutual orientation between these frames is defined by the three orientation angles $A_1$, $A_2$, and $A_3$ around the three ICRS Cartesian axes.
Since the catalogs under consideration are close to each other at sub-arcsecond level, the orientation of a vector $(X,Y,Z)$, can be written in the following simple form:

\begin{equation}
\left( \begin{array}{ccc} X_1 \\ Y_1 \\ Z_1 \end{array} \right) = 
\left( \begin{array}{ccc} 1 & A_3 & -A_2 \\ -A_3 & 1 & A_1 \\ A_2 & -A_1 & 1 \end{array} \right)
\left( \begin{array}{ccc} X_2 \\ Y_2 \\ Z_2 \end{array} \right) \,.
\label{eq:orientation_xyz}
\end{equation}

Taking into account that on the celestial sphere
\begin{equation}
\left( \begin{array}{ccc} X \\ Y \\ Z \end{array} \right) = 
\left( \begin{array}{ccc} \cos\alpha\cos\delta \\ \sin\alpha\cos\delta \\ \sin\delta \end{array} \right) \,,
\label{eq:r_vector}
\end{equation}
and turning to the differences between the object positions in two catalogs $\Delta\alpha=\alpha_1-\alpha_2$ and $\Delta\delta=\delta_1-\delta_2$,
we derive the final expression:
\begin{equation}
\begin{array}{rcl}
\Delta\alpha & = & \phantom{-}A_1\cos\alpha\tan\delta + A_2\sin\alpha\tan\delta - A_3 \,, \\[1ex]
\Delta\delta & = & -A_1\sin\alpha + A_2\cos\alpha \,.
\end{array}
\label{eq:orientation_ad}
\end{equation}

The system of Eqs.~\ref{eq:orientation_ad} for all or selected common objects in two catalogs is solved by the least squares method (LSM) to determine
the orientation angles $A_1$, $A_2$, and $A_3$ between two CRF realizations and their errors (uncertainties).

Generally speaking, the differences between the two catalogs include not only the rotational part but also other, mostly coordinate-dependent terms
that describe the systematic errors in the compared catalogs, including distortion of the celestial frames realized by the catalogs.
Determination of the systematic errors of the celestial object positions in catalogs is a traditional and well developed astrometric task,
see, e.g., \citet{Sokolova2007} and papers cited therein.
Since these systematic errors might influence the orientation angles, they should be estimated during the GCRF--ICRF alignment procedure.


\section{Overview of recent activity}
\label{sect:overview}

The GCRF astrometric catalog will join both galactic stars and extragalactic objects in a single highly-accurate system.
As a next stage, this catalog will be aligned to the ICRS using common GCRF and ICRF objects.
Two tasks should be solved to provide such an alignment, analogously to what was done for the {\it Hipparcos} catalog
\citep{Lindegren1995,Kovalevsky1997}:
\begin{enumerate}
\item Determination of mutual orientation between the two frames at an initial epoch, most naturally at the {\it Gaia} mean observation epoch, which
is expected to be $\sim$2017.0 or somewhat later if the mission will be prolonged after its 5-year initially planned duration.
\item Determination of the mutual rotation between two systems.
\end{enumerate}

For the {\it Hipparcos} catalog, the achieved accuracy was 0.6~mas at the epoch 1991.25 for orientation and 0.25~mas/yr for rotation \citep{Kovalevsky1997}.
In the case of the GCRF, the desired accuracy is about an order better, which is quite a challenge.

On the outset, the accuracy of the link between GCRF and ICRF will depend on the number of common objects and their astrometric quality, i.e. accuracy of their coordinates in two catalogs.
If Gaia just observes all the objects it can detect with nearly uniform accuracy over the sky depending mainly on the object's optical brightness,
the situation with the ICRF sources is more complicated.
The ICRF2 catalog is very inhomogeneous in respect to the radio source position errors due to the large difference in number of observations
from 3 to 337,322, see Section~\ref{sect:radio_frame} for more detail.
The situation has improved substantially with realization of the project on re-observation of the VCS (VLBA Calibrator Survey) sources \citep{Gordon2016},
which allowed a significant improvement in the accuracy of about 2000 ICRF source positions (see Section~\ref{sect:radio_frame}), but it is still far from ideal. 

Current activities in preparation towards aligning the GCRF with the ICRF are developing in several directions:
\begin{enumerate}
\item Selection of prospective link radio sources and their intensive observations.
\item Photometric optical observations of radio sources.
\item Optical astrometric observations of radio sources with ground-based telescopes.
\item Creation of data banks of optical images of ICRF objects.
\end{enumerate} 

Several years ago, \citet{Bourda2008} started work on selecting optically bright radio sources of good radio astrometric quality, i.e. having sufficient flux density and compact structure.
Finally, 195 prospective link sources were selected.
The various stages of this work were described by \citet{Bourda2008} (selection of optically bright radio sources),
\citet{Bourda2010} (source detection on the long VLBI baselines), \citet{Bourda2011} (source imaging to estimate their radioastrometric quality),
\citet{LeBail2016} (improving radio positions of selected sources).
The program is being continued.

\citet{Zacharias2014} obtained accurate optical positions of 413 AGN, and investigated their errors and their impact on the accuracy of the link between radio and optical frames. Comparison of the optical positions with radio positions showed that the differences statistically exceed the known errors in the observations.
The physical offset between the optical and radio emission centers was identified as a likely cause.
This effect, called by the authors detrimental, astrophysical, random noise (DARN), was found to be at $\sim$10~mas level.
The authors came to the conclusion that the GCRF--ICRF orientation angles can hardly be determined with an error better than 0.5~mas, without a substantial increase in the number of the link objects.
This estimate was based on ground-based results and can be somewhat improved with Gaia observations, but the DARN can prevail in this case too.

The new catalog URAT1 contains positions of over 228 million objects in the magnitude range of about $R$=3--18.5 with a typical error of 10--30~mas
and proper motions of over 188 million objects with a typical error of 5--7~mas/yr \citep{Zacharias2015}.
There are two shortcomings of this catalog limiting its usefulness for studies on the GCRF--ICRF link.
Firstly, the catalog covers only just over a half of the sky, namely the region with $\delta \ge 15^{\circ}$.
Secondly, most of the astrometric radio sources are absent in the URAT1 catalog as they are fainter than 18.5$^m$ (see Fig.~\ref{fig:mag_n}). 
However, this work is of great importance as it provides very valuable information on the shift between radio and optical positions for hundreds of radio sources.

A dedicated program of photometric observations of the ICRF sources have been conducted under coordination of the Paris Observatory
\cite{Taris2013,Taris2015,Taris2016}.
These observations provide much new information about the optical brightness of the astrometric radio sources as well as about their optical variability.
The results reported by this group showed that the peak-to-peak change in optical magnitude is typically at a level of several tenths of magnitude,
and exceed 1$^m$ for many objects reaching sometimes 3$^m$, e.g., for the source B0716+714.

\citet{Andrei2012,Andrei2014} have been working on the compilation of the {\it Gaia} QSO catalog.
Data from different surveys and catalogs are assembled to complete QSO characteristics including positions, photometry, morphology, and imaging.
This work, in particular, provides useful data for investigation of the radio minus optics position shift.
The most complete bank of optical images of the ICRF objects was created by \citet{Andrei2015}.

\citet{Souchay2015} presented the 3rd release of the Large Quasar Astrometric Catalog (LQAC-3).
This catalog contains accurate positions, magnitudes in 9 bands $UBVGRIZJK$, radio flux in 5 bands from 750~MHz to 30~GHz, morphology indexes
in $BRI$ bands, and absolute magnitude in $B$ and $R$ band for 321,957 objects, primarily quasars, including $\sim$5\% of other AGN types. 

All these works are directed, in particular, towards better link source selection and the improvement in the accuracy of their radio astrometric positions,
which provides better accuracy of the GCRF--ICRF link.
In the next sections, new possibilities to improve the quality of the link will be discussed.


\section{Radio frame}
\label{sect:radio_frame}

All the recent studies discussed in Section~\ref{sect:overview} are based on using the ICRF2 as a radio CRF.
However, the ICRF2 catalog created in 2009 \citep{Fey2015} is already outdated.
It is expected that the first link between the GCRF and ICRF will be performed during 2018--2019 using the next VLBI-based ICRF realization, ICRF3 \citep{Jacobs2014}.
Currently, the radio source position catalog gsf2015b\footnote{http://gemini.gsfc.nasa.gov/solutions/astro/}
derived at the NASA Goddard Space Flight Center (GSFC) VLBI Group appears to be the closest to the future ICRF3.
It is computed using about the same data analysis strategy as was used for computation of ICRF2 (also at GSFC) but involves many more observations (VLBI delays). 
Table~\ref{tab:icrf2-gsf} contains a statistical comparison between the gsf2015b and ICRF2 catalogs.

\begin{table}
\centering
\caption{Basic statistics of the ICRF2 and gsf2015b catalogs.}
\label{tab:icrf2-gsf}
\begin{tabular}{lccrccc}
\hline
\multicolumn{1}{c}{Catalog} & Sources & Sessions & \multicolumn{1}{c}{Delays} & Period & \multicolumn{2}{c}{Median error, $\mu$as} \\
         &      &      &            &     & $\alpha \, \cos\delta$ & $\delta$ \\
\hline
ICRF2    & 3414 & 4540 &  6,495,553 & 1979.08.03 -- 2009.03.16 & 397 & 739 \\
gsf2015b & 4089 & 5836 & 10,453,527 & 1979.08.03 -- 2015.11.09 & 123 & 210 \\
\hline
\end{tabular}
\end{table}

\begin{figure}
\begin{center}
\includegraphics[width=\textwidth]{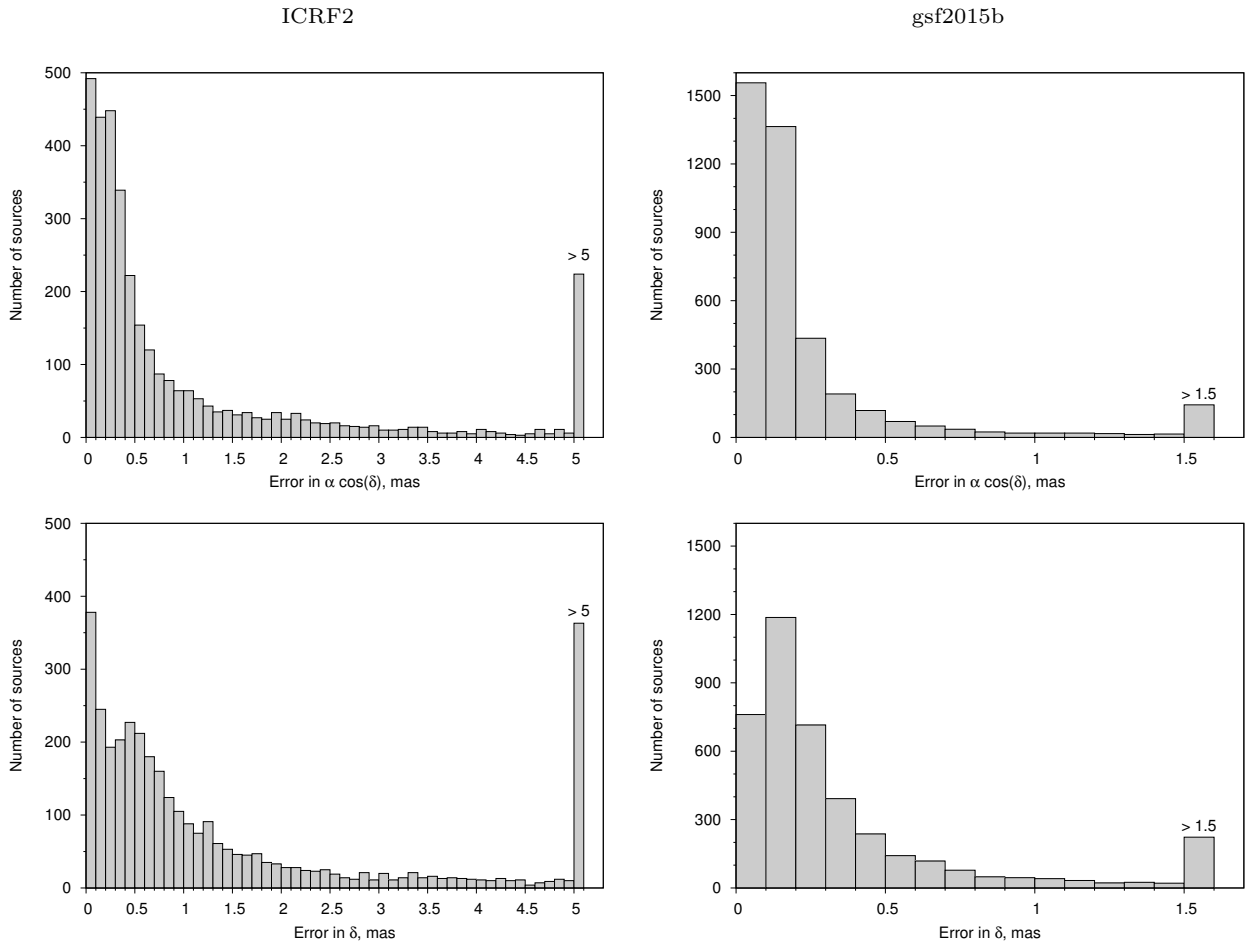}
\caption{Distribution of the source position errors in ICRF2 (left) and gsf2015b (right) catalogs.}
\label{fig:gsf-icrf2_errors}
\end{center}
\end{figure}

Figure~\ref{fig:gsf-icrf2_errors} illustrates the position uncertainty distribution in two catalogs.
One can see that the overall level of the position uncertainty in the gsf2015b catalog is much smaller than that in the ICRF2 catalog.

\begin{figure}
\begin{center}
\includegraphics[width=\textwidth]{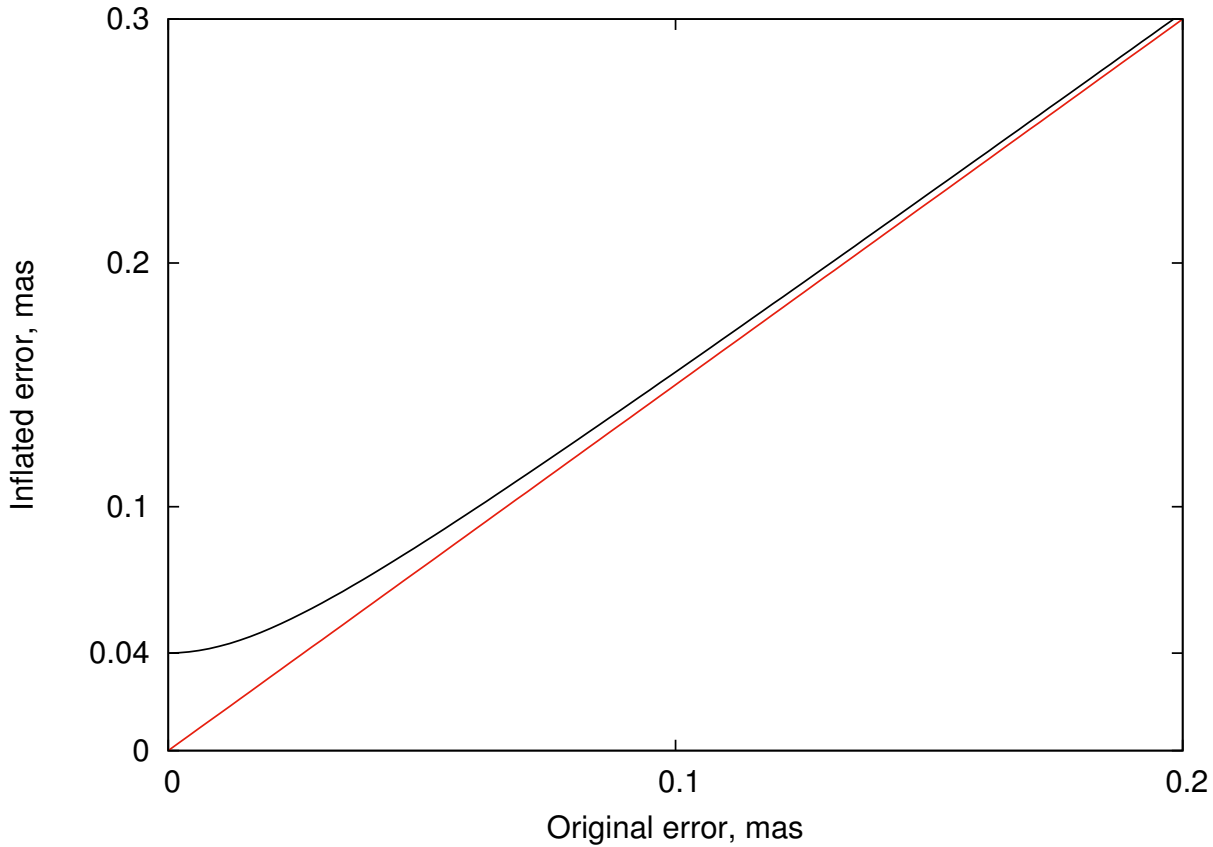}
\caption{Dependence of the inflated ICRF2 position errors on the original ones. Black line corresponds to the inflation formula
  $\sigma^2_{inflated} = (1.5 \, \sigma_{computed})^2 + (0.04~\mathrm{mas})^2$ used in \citet{Fey2015}. Red line corresponds to simple
  regression $\sigma_{inflated} = 1.5 \, \sigma_{computed}$, i.e. to the ICRF2 formula with the error floor omitted.}
\label{fig:error_inflation}
\end{center}
\end{figure}

It can be seen from Table~\ref{tab:icrf2-gsf} and Fig.~\ref{fig:gsf-icrf2_errors} that the GSFC catalog is more advanced when compared with the ICRF2.
It should be noted that the gsf2015b catalog provides the original position errors obtained from the LSM solution, while ICRF2 position errors
were inflated following the formula $\sigma^2_{inflated} = (1.5 \, \sigma_{computed})^2 + (0.04~\mathrm{mas})^2$ \citep{Fey2015},
see Fig.~\ref{fig:error_inflation}.
The error floor of 0.04~mas is mostly effective for small original errors less than 0.1--0.2~mas; for larger original
errors inflated errors may be taken as original errors multiplied by factor 1.5. 
Nevertheless, even taking this factor into account, gsf2015b source positions are substantially more precise than those of the ICRF2.
The primary reason for the significant improvement in source position errors is re-observation of about
2000 VCS (VLBA Calibrator Survey) sources \citep{Gordon2016}, which is approximately 2/3 of the total number of the ICRF2 sources.

Based on these considerations, the gsf2015b catalog was used in this study as a prototype of ICRF3.

\begin{figure}
\begin{center}
\includegraphics[width=\textwidth]{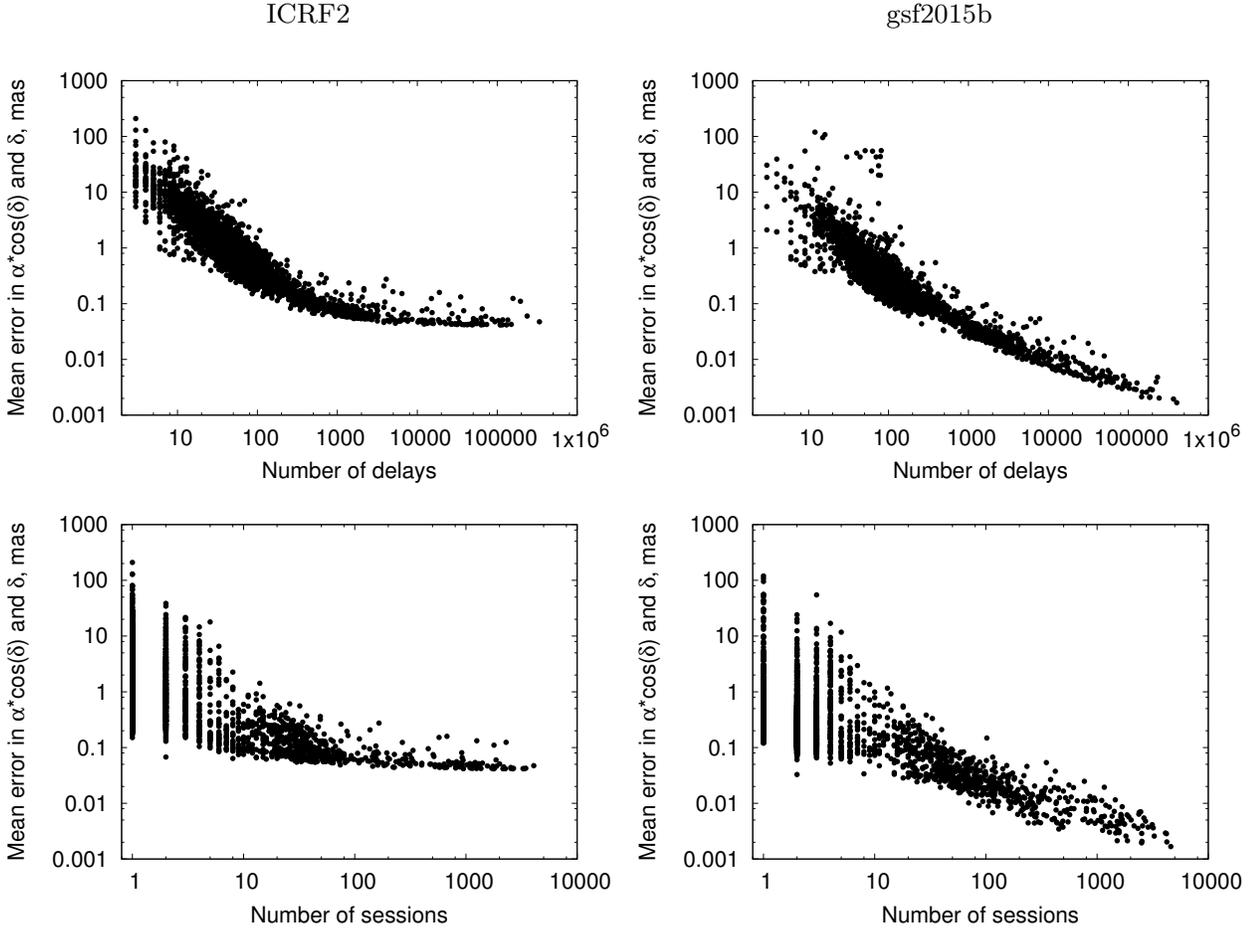}
\caption{Dependence of source position errors on the number of observations for the ICRF2 (left) and gsf2015b (right) catalogs.}
\label{fig:gsf-icrf2_errors_nobs}
\end{center}
\end{figure}

One of the permanent tasks of the VLBI community is improving accuracy of the radio source positions.
Due to limited VLBI resources, proper planning of the observations is important.
Figure~\ref{fig:gsf-icrf2_errors_nobs} illustrates the dependence of the source position uncertainty on the number of observations (delays and sessions)
for the ICRF2 and gsf2015b catalogs.
The error floor of 0.04~mas introduced in the final catalog \citep{Fey2015} can clearly be seen in the plot for ICRF2.
This analysis shows that the source position uncertainties depend primarily on the number of delays, and to lesser extent on the number of sessions.
It also suggests that to reliably achieve sub-mas position error, about 100 observations (VLBI delays) should be obtained.

\section{Increasing the number of link radio sources}
\label{sect:more_sources}

A criterion for the initial source selection in \citet{Bourda2008}, which was the base for the consequent works \citet{Bourda2010,Bourda2011,LeBail2016},
was using ICRF2 sources with optical magnitude $\le 18^m$.
The latter limit was defined to use the objects with a small {\it Gaia} position error $<\sim$70~$\mu$as as was estimated during {\it Gaia} pre-launch analysis \citep{Lindegren2008}. To select the optically bright radio sources, the catalog of \citet{Veron2006} was used as the source of the photometric data.

It seems that the approach applied by \citet{Bourda2008} for the link source selection can be substantially improved in view of new data that became available during recent years.
First, as was discussed in Section~\ref{sect:radio_frame}, the ICRF2 does not provide the best choice of astrometric radio sources with reliable highly-accurate coordinates.
Second, the catalog \citet{Veron2006} does not contain sufficiently complete photometric data as compared with the latest catalogs. 
So, an improved strategy for preliminary link source selection can be suggested.
A new approach can include the use of the latest VLBI-based CRF solutions containing more radio sources with accurate positions, especially
in the Southern Hemisphere, and the catalog OCARS (Optical Characteristics of Astrometric Radio Sources, \citet{Malkin2016d})
that contains the most complete photometric data for astrometric radio sources and thus provides the maximum choice for selection
of optically bright radio sources.

The second option that is worth investigating is using radio sources with a magnitude $18<G\le20^m$, where $G=20^m$ is the threshold for $Gaia$
observations \citep{Perryman2001,Lindegren2008}.
Although fainter sources will have much larger {\it Gaia} positional error, the number of these sources may compensate for such a loss of precision.
An analysis based on the rotation covariance matrix analysis was performed by \citet{Mignard2014}.
It was found that moving from $18^m$ to $20^m$ threshold leads to about a twofold increase of ICRF2 link sources from $\sim$500 to $\sim$1000,
and reduction of the errors of the orientation angles from $\sim$7~$\mu$as to $\sim$6~$\mu$as.

\begin{figure}
\begin{center}
\includegraphics[width=\textwidth]{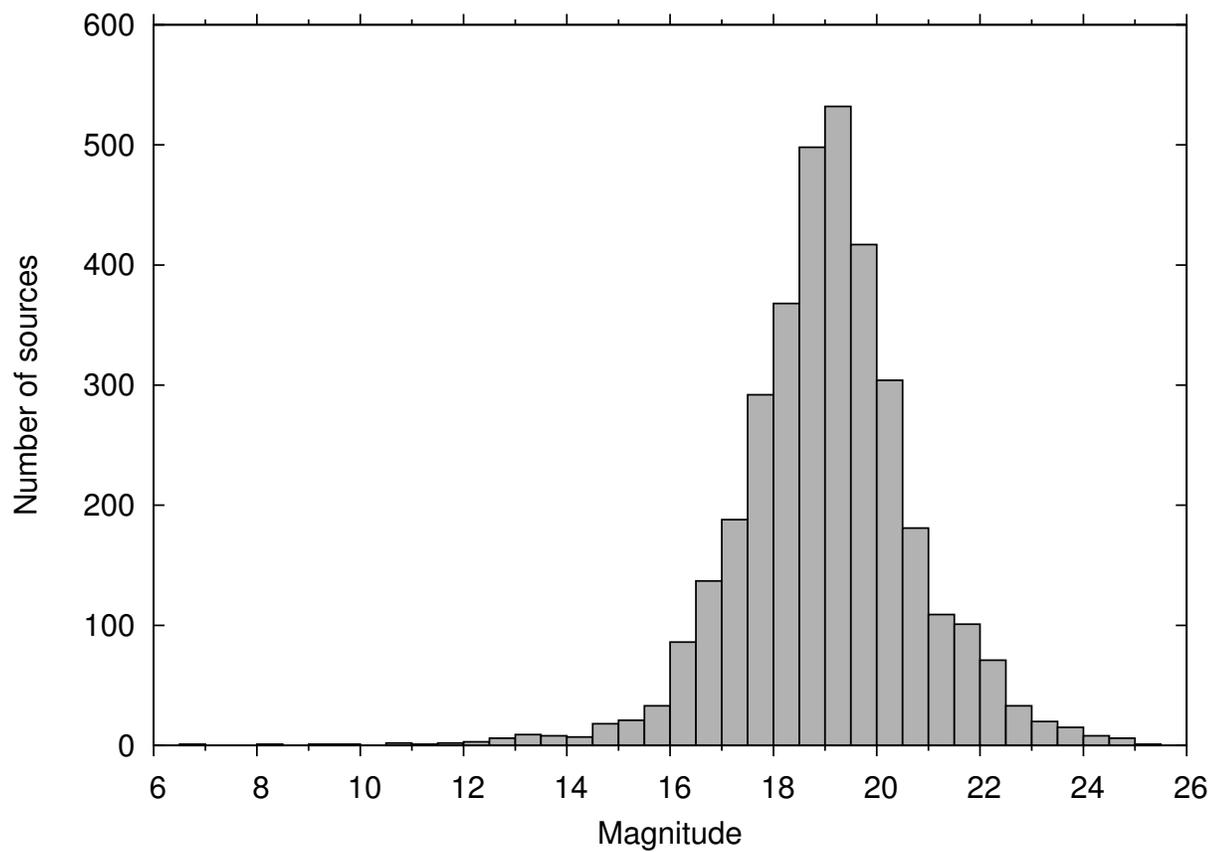}
\caption{Distribution of optical magnitudes in the gsf2015b catalog.}
\label{fig:mag_n}
\end{center}
\end{figure}

Another approach based on Monte Carlo simulation was used in the current study.
A list of the link sources for these computations was composed of all AGN with the optical magnitude $\le 20^m$ and galaxies with the optical
magnitude $\ge 16^m$ selected from the gsf2015b catalog.
It is supposed that faint galaxies can be expected to be unresolved objects for {\it Gaia} \citep{deSouza2014}.
Optical magnitudes for these sources were taken from the OCARS catalog \citep{Malkin2016d}.
Figure~\ref{fig:mag_n} illustrates the distribution of the optical magnitudes in the source set used in analysis.

\begin{figure}
\begin{center}
\includegraphics[width=\textwidth]{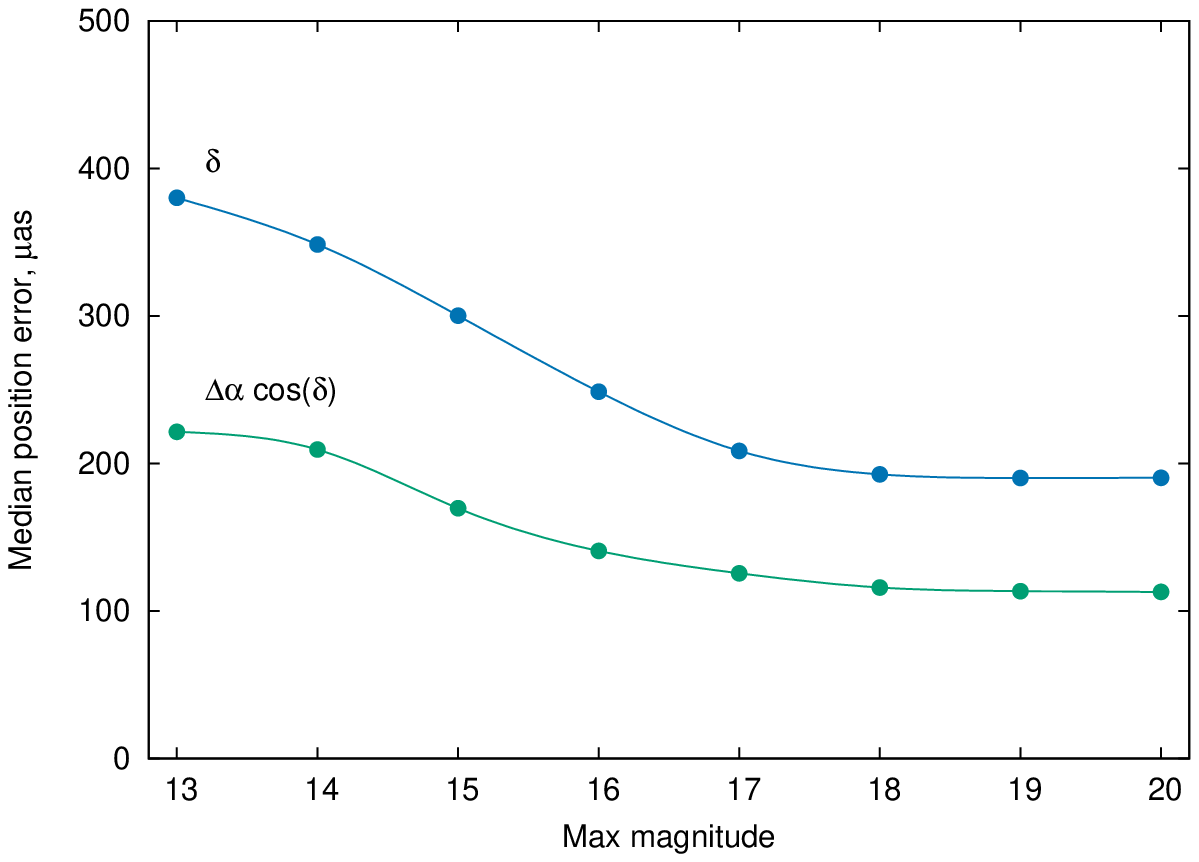}
\caption{Median errors of the gsf2015b positions depending on source selection based on the faint-end optical magnitude threshold.}
\label{fig:gsf_median_err}
\end{center}
\end{figure}

The median uncertainty of the gsf2015b source positions for different sets of sources selected with different upper limits for the optical magnitude
is depicted in Fig.~\ref{fig:gsf_median_err}.
One can see that moving to the optically fainter sources improves the overall position precision.

\begin{figure}
\begin{center}
\includegraphics[width=\textwidth]{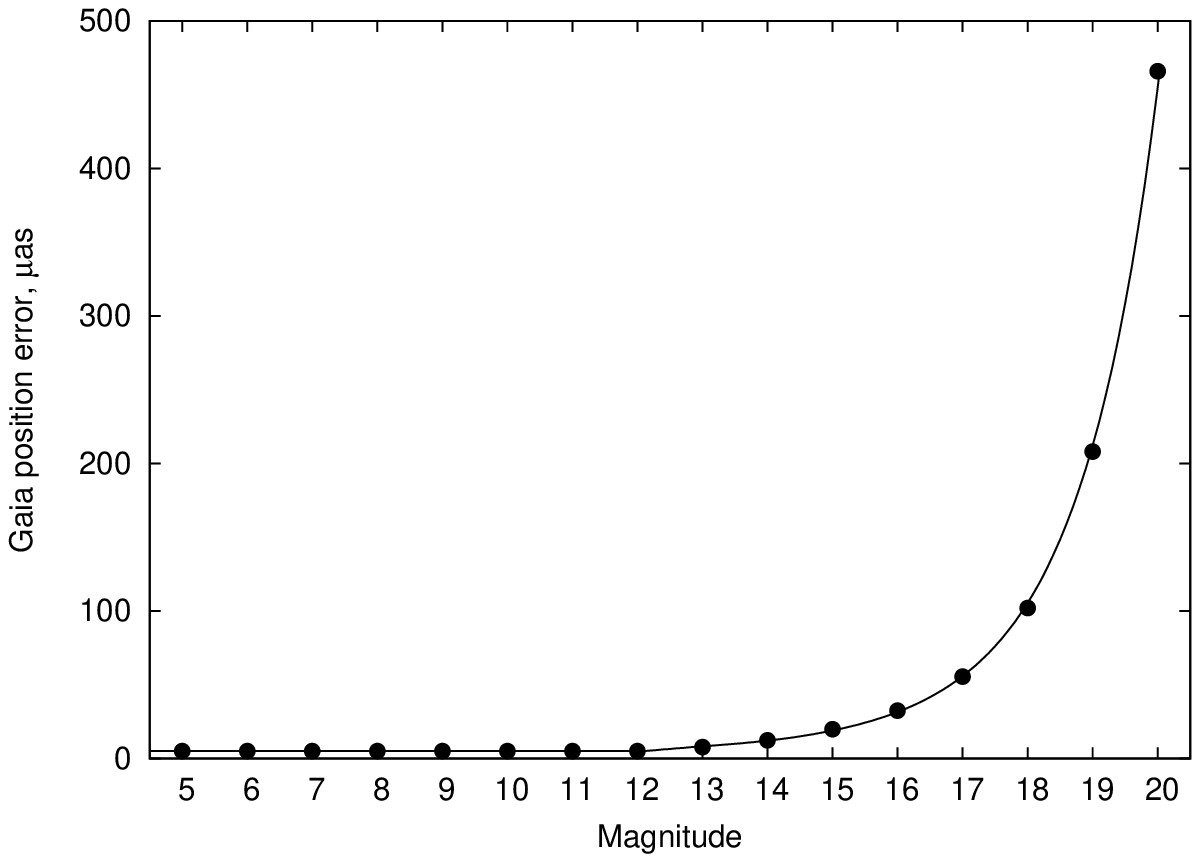}
\caption{Expected {\it Gaia} position error according to \citet{Bruijne2014} (points).
Solid line corresponds to the approximation function (\ref{eq:gaia_error}).}
\label{fig:gaia_error}
\end{center}
\end{figure}

For the VLBI position errors, the uncertainty estimates given in the gsf2015b catalog were used.
The {\it Gaia} position errors used during Monte Carlo simulation were estimated in the following way.
Expected {\it Gaia} parallax standard error $\sigma_\pi$ depending on the optical brightness of the object $G$ is given by \citet{Bruijne2014}
(points in Fig.~\ref{fig:gaia_error}).
The authors also provide a rather complicated formula representing $\sigma_\pi$ depending on the optical brightness.
However, this formula appears to be unnecessarily complicated for the simulation studies, it depends not only on the $G$ magnitude but also on the $V-I$ color,
which is not known for most of the astrometric radio sources.
So, a simpler approximation of the $\sigma_\pi$ was derived (in $\mu$as):

\begin{equation}
\sigma_\pi = \left\{\begin{array}{ll}
6.7,                                               & G < 12.1\,, \\[1ex]
6.7 + 4.86 \, (G-12.1),                            & 12.1 \le G < 13 \,, \\[1ex]
10^{(1.044+0.1528 \, (G-13)+0.01373 \, (G-13)^2)}, & G \ge 13 \,.
\end{array}
\right.
\label{eq:gaia_error}
\end{equation}

Then the {\it Gaia} position error is computed as $\sigma_0 = 0.743 \, \sigma_\pi$ \citep{Bruijne2014}.
This approximation function is depicted in Fig.~\ref{fig:gaia_error}.

\begin{figure}
\begin{center}
\includegraphics[width=\textwidth]{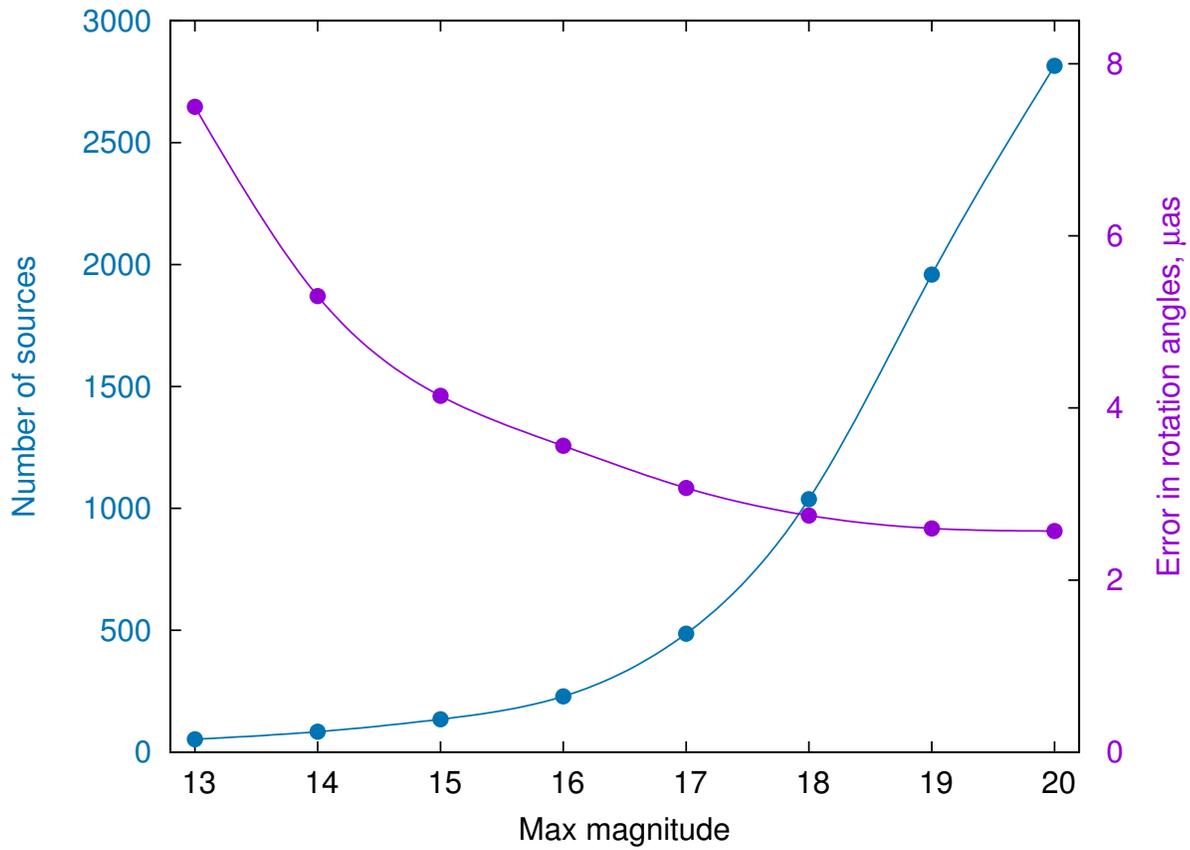}
\caption{Simulated error in orientation angles depending on source selection based on the faint-end optical magnitude threshold.}
\label{fig:err_mag}
\end{center}
\end{figure}

The result obtained with Monte Carlo simulation (10,000 iterations) is shown in Fig.~\ref{fig:err_mag}.
This result is generally similar to that obtained by \citet{Mignard2014} using a different method, however the errors
in the orientation angles are about two times smaller in the current work, in particular, due to larger number of sources used.

\begin{figure}
\begin{center}
\includegraphics[width=\textwidth]{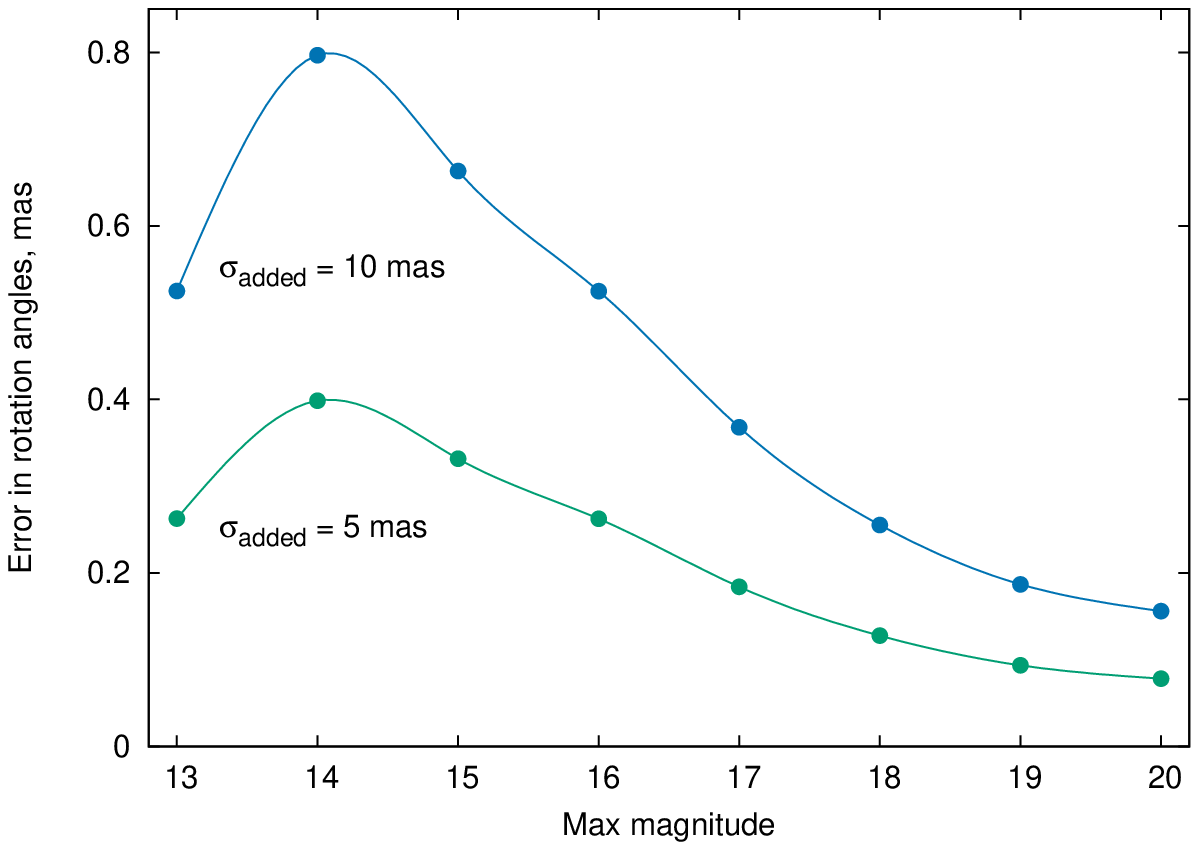}
\caption{Simulated error in orientation angles depending on the additive error due to radio--optics position shift.}
\label{fig:err_mag_add}
\end{center}
\end{figure}

Surely, both results are too optimistic because they were obtained without taking into account the radio--optics position shift,
such as the DARN mentioned in Section~\ref{sect:overview}.
To estimate the impact of this effect, an additive error $\sigma_{\mathrm{added}}$ was added in quadrature to the simulated ICRF--GCRF coordinate differences.
The result of this test obtained by Monte Carlo simulation in the same way as the previous one is presented in Fig.~\ref{fig:err_mag_add}.
The obtained error in the orientation angles agrees with the value of 0.5~mas predicted by \citep{Zacharias2014} for $\sigma_{\mathrm{added}}$ = 10~mas.

\begin{figure}
\begin{center}
\includegraphics[width=\textwidth]{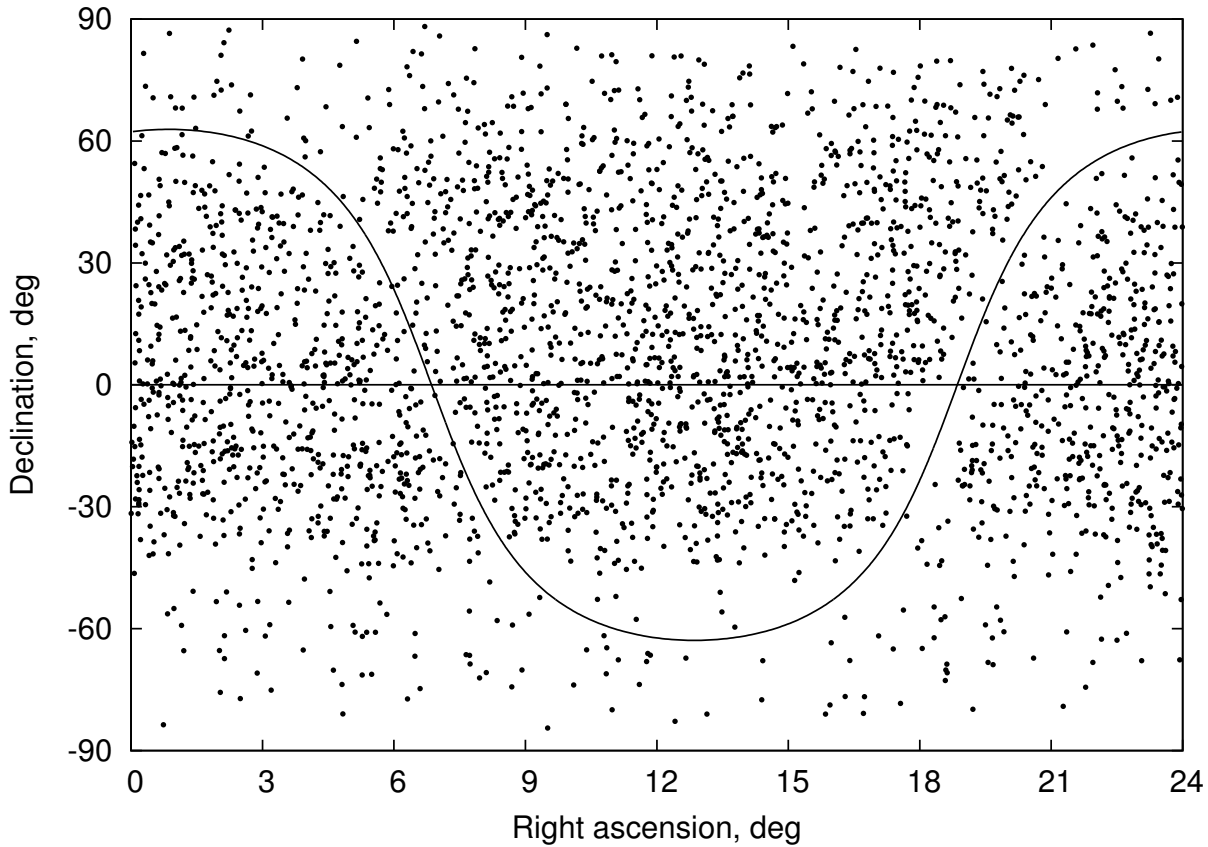}
\caption{Distribution of the gsf2015b sources with optical magnitude $\le 20^m$ over the sky.}
\label{fig:gsf20_sky}
\end{center}
\end{figure}

Finally, all the results presented in this section showed that it is advisable to use all compact radio sources up to 20$^m$ for the ICRF--GCRF link.
The distribution of 2815 gsf2015b sources with the optical magnitude $\le 20^m$ over the sky is shown in Fig.~\ref{fig:gsf20_sky}.
Indeed, this is the first stage of selection of the best ICRF--GCRF link sources that should follow by estimation of their radioastrometric quality
as discussed in \citet{Bourda2008}.
However, it allows one to have several times more prospective link sources at this stage than was considered earlier: 2815 sources versus
535 sources selected by \citet{Bourda2008}.

\section{Using radio stars}
\label{sect:radio_stars}

The problem of aligning the GCRF to ICRF is similar to the problem of aligning the HCRF to the ICRF.
One of the methods applied for this purpose during the {\it Hipparcos} mission was using radio stars.
It proved to be the most accurate method among others considered for the orientation of the {\it Hipparcos} catalog \citep{Kovalevsky1997}.
The error in the orientation angles between the two frames obtained from VLBI observations of 12 radio stars was estimated to be about 0.5~mas.

Later, \citet{Boboltz2007} reported on the results of observations of 46 radio stars with flux density of 1--10~mJy obtained with the
VLA (Very Large Array) plus the Pie Town antenna of the VLBA (Very Long Baseline Array) narrow regional network.
The position of radio stars were determined by means of phase referencing to close ICRF sources with an error of about 10~mas on average.
The HCRF--ICRF orientation angles were estimated with an error of $\sim$2.7~mas.

The current accuracy of both VLBI and optical ({\it Gaia}) observations is much better.
Consequently, properly scheduled radio star observations using large regional or global VLBI networks can provide much smaller position errors 
and hence better accuracy of the link between optical and radio frames.
A Monte Carlo simulation was performed by \citet{Malkin2016e} to estimate an error in the orientation angle between GCRF and ICRF obtained
from radio stars observation.
Results of this study showed that VLBI observations of radio stars can provide an independent and accurate method to link the GCRF to the ICRF.
A properly organized VLBI program for radio star observations can lead to the realization of the GCRF--ICRF link with an error of about 0.1~mas with a reasonable load on the VLBI network.
Thus, this method can provide a valuable contribution to the improvement of the GCRF--ICRF link.

Details of this work are given in \citet{Malkin2016e}.


\section{Galactic aberration}
\label{sect:ga}

Comparison of the ICRF and GCRF catalogs should be made at a certain epoch $t_0$.
Two natural choices are the current standard epoch of the astronomical equations and quantities $t_0$=J2000.0, and the mean epoch
of the {\it Gaia} catalog $t_0$=$\sim$2017.0, supposing a 5-year period of {\it Gaia} operations starting from July 2014.
It can be reasonably supposed that the {\it Gaia} positions will be brought to $t_0$ using its own proper motions.

The situation with the ICRF positions is not so simple.
The ICRF concept is based on the absence of detectable motions of ICRF sources.
It is definitely not the case at the micro-arcsecond level of accuracy.
Several works showed that astrometric VLBI observations are capable of revealing statistically meaningful apparent motions of the ICRF objects,
see, e.g., \citet{MacMillan2005,Titov2013}.
The question of the nature of these motions is very complicated, and the consistency between various estimates is not satisfactory yet. 
The overall problem of source motions is beyond the scope of this paper.
Here, only one systematic component of source motion pattern, namely galactic aberration in proper motions (GA) is discussed.
The theory of this effect is considered in \cite{Kovalevsky2003,Kopeikin2006,Liu2012,Liu2013}.
The proper motion caused by the GA is given by \citep{Malkin2011fe}:
\begin{equation}
\begin{array}{rcl}
\mu_l \cos b &=& -A \sin l \,, \\
\mu_b &=& -A \cos l \sin b \,, \\
\end{array}
\label{eq:galactic}
\end{equation}
where $l$ and $b$ are the galactic longitude and latitude of the object, respectively, and $A$ is the GA constant depending on the galactic rotation
parameters.
The most probable value of the GA constant is $A = 5 \pm 0.3~\mu$as/yr \citep{Malkin2014b}.

To estimate the impact of the GA on the orientation angles between ICRF and GCRF, a special test was performed.
The idea of this test was proposed by \citet{Malkin2015c}and is extended here.
Two catalogs of radio source positions were used for the simulation.
The first catalog comprises 688 ICRF2 sources of AGN type and with visual magnitude 18$^m$ or brighter following the principles of the
link source selection proposed by \citet{Bourda2008}.
The second catalog consists of 2815 gsf2015b sources discussed in Section~\ref{sect:radio_frame}.

All the existing catalogs of radio source positions are derived without accounting for the GA during data processing.
Therefore, to bring the selected link source positions to the epoch $t_0$ the following equations should be used:
\begin{equation}
\begin{array}{rcl}
\alpha(t_0) & = & \alpha(t) - \mu_\alpha (t-t_0) \,, \\
\delta(t_0) & = & \delta(t) - \mu_\delta (t-t_0) \,, \\
\end{array}
\label{eq:2epoch}
\end{equation}
where $t$ is the mean epoch of observations of the source in the ICRF2 or GSFC catalogs,
$\alpha(t)$ and $\delta(t)$ are source coordinates (right ascension and declination, respectively) in the catalog,
$\mu_\alpha$ and $\mu_\delta$ are GA-induced motions in right ascension and declination, respectively,
computed by Eq.~\ref{eq:galactic} and transformed to the equatorial system as described by \citet{Malkin2014b,Malkin2015d}.
All the computations were made for $t_0$=J2000.0 and $t_0$=2017.0.

Then, we have two catalogs for each of four variants (two initial catalogs and two $t_0$ epochs).
The first catalog in each pair is merely the initial catalog.
Such a catalog would be used for the ICRF--GCRF link if GA is not taken into account, which is currently the case.
The second catalog contains positions of the same sources transferred to the epoch $t_0$ for the GA-induced proper motions.
This catalog would correspond to the radio source positions computed with taking into account the GA effect during VLBI data processing.
Consequently the orientation angles between the two catalogs were computed.
Results are presented in Table~\ref{tab:ga_orientation}.

\begin{table}
\centering
\caption{Impact of galactic aberration on the orientation angles between the ICRF and GCRF. Unit: $\mu$as.}
\label{tab:ga_orientation}
\begin{tabular}{lcccc}
\hline
\multicolumn{1}{c}{Catalog} & $t_0$ & $A_1$ & $A_2$ & $A_3$ \\
\hline
ICRF2    & J2000.0 & $  ~~~1.3 \pm 0.5 $ & $  -0.1 \pm 0.5 $ & $ 0.3 \pm 0.4 $ \\
         & 2017.0  & $  ~~26.8 \pm 2.4 $ & $  -4.9 \pm 2.3 $ & $ 4.0 \pm 2.0 $  \\
gsf2015b & J2000.0 & $   -19.4 \pm 0.5 $ & $ ~~1.9 \pm 0.5 $ & $ 0.6 \pm 0.4 $  \\
         & 2017.0  & $  ~~36.1 \pm 0.6 $ & $  -4.2 \pm 0.6 $ & $ 2.7 \pm 0.6 $ \\
\hline
\end{tabular}
\end{table}

\begin{figure}
\begin{center}
\includegraphics[width=\textwidth]{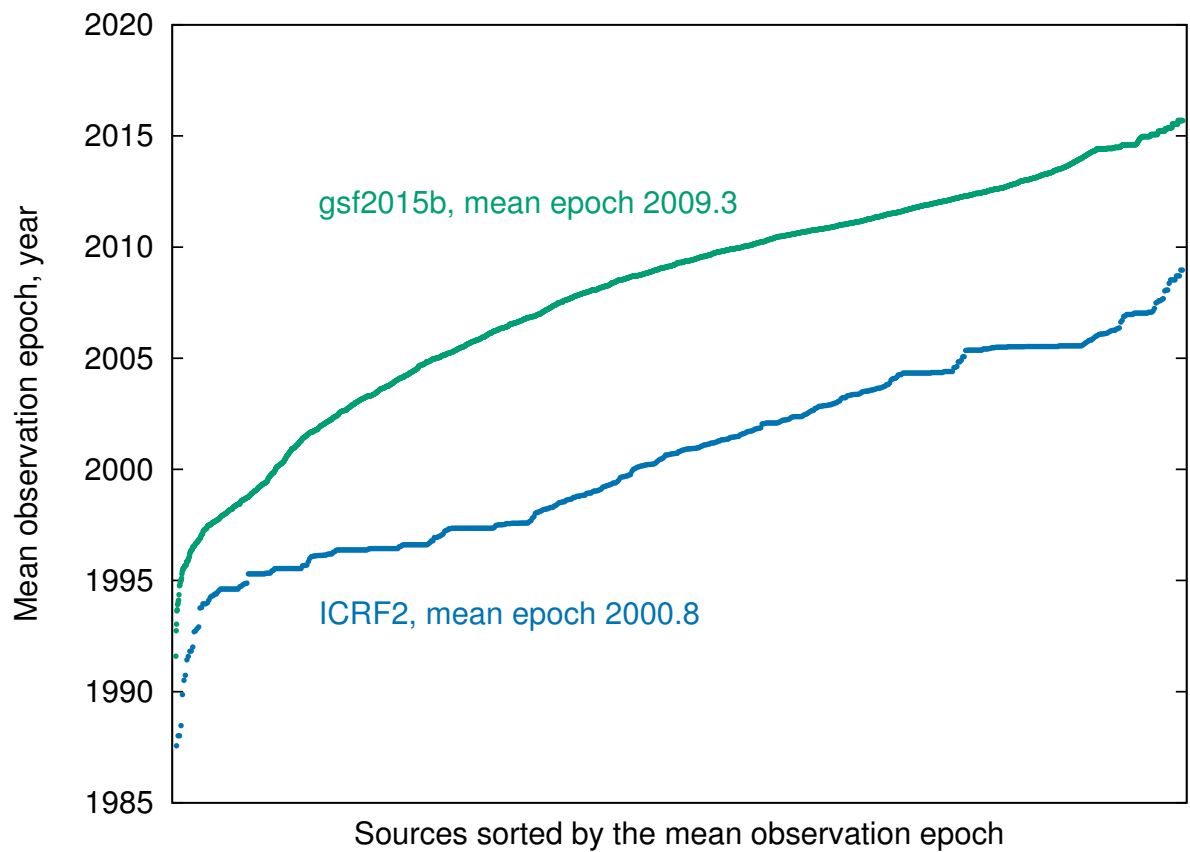}
\caption{Mean source observations epoch for the ICRF2 and gsf2015b catalogs.}
\label{fig:cat_epochs}
\end{center}
\end{figure}

The first line of Table~\ref{tab:ga_orientation} corresponds to computations made in \citet{Malkin2015c}.
The three other lines are the test extension that allowed us to correct the earlier preliminary conclusion.
After the preliminary test (first line) a conclusion was drawn that the impact of the GA on the ICRF--GCRF link is practically negligible.
More detailed test performed in this study has shown that the GA effect can be substantial.
This can be seen from both the significant value of the orientation angles and their uncertainties.
The errors in the orientation angles depend on both the number of common sources in the catalogs and the time interval between the mean source
observation epoch in the catalog (let us call it catalog epoch) and $t_0$.
Mean epochs for two catalogs are shown in Fig.~\ref{fig:cat_epochs}. 
The ICRF2 catalog epoch is closer to J2000.0 than the gsf2015b epoch, but the latter contains 4 times more sources, which resulted in about
the same errors in the orientation angles for $t0$=J2000.0.
In contrast, the gsf2015b catalog epoch is closer to 2017.0, which together with larger number of sources gives much smaller errors in
the orientation angles as compared with the ICRF2.

\section{Discussion and conclusion}

A great milestone in the construction of the celestial reference frame is the {\it Gaia} mission, which will result in the GCRF with expected accuracy of a few tens of micro-arcsecond for final release in the early 2020s.
A new ICRF release based on VLBI observations of extragalactic radio sources of similar accuracy can be also expected by that time.
Constructing of a single multi-frequency celestial reference frame based on the ICRF and GCRF is the primary task of fundamental astronomy
for the next decade \citep{Gaume2015}.

The first step toward this goal is an accurate alignment of the {\it Gaia} catalog to the ICRF. The desired goal is to achieve mutual orientation between the two frames with an accuracy of 0.1~mas or better, which is a challenging task.

Currently, the International VLBI Service for Geodesy and Astrometry (IVS) is conducting a special program on observation of prospective 
link sources not having a sufficient number of observations in the framework of regular observing programs \citep{LeBail2016}.
This forms part of the plan to prepare to achieve the GCRF--ICRF link using 195 selected sources selected on basis of several criteria, such as,
inclusion in the ICRF2, optical magnitude $\le 18^m$, symmetric compact structure, sufficient radio flux density.
The two first criteria seem to be outdated.
Firstly, the ICRF2 is currently not the most appropriate radio source position catalog that can serve as an ICRF3 prototype, which is planned to be used
for initial alignment of the {\it Gaia} catalog to the ICRS as discussed in Section~\ref{sect:radio_frame}.
Secondly, as shown in Section~\ref{sect:more_sources}, using optically fainter sources up to 20$^m$ provides more precise determination
of the orientation angles.
It is even more important that the use of more sources with reliable VLBI positions is  necessary to mitigate the impact of most of the negative
factors mentioned below, see Fig.~\ref{fig:err_mag_add} and related text.
As discussed, there are possibilities to multiply the number of link sources taking into account the substantial increase of the number
of radio sources with reliably determined positions and the number of sources having photometry measurements.

The following problems were identified in the literature that dilute the accuracy of the link between radio and optical frames:
\begin{itemize}
\item Structure effects (discussed below).
\item Systematic errors of radio position catalogs.
\item Multiple radio sources related to a single object in optics, e.g., binary black holes, and vice versa.
\item Gravitational lenses.
\item Errors in ICRF--GCRF cross-identification.
\end{itemize}

Most probably, the main factor that will deteriorate the accuracy of the link between the radio and optical frames, is the source structure.
It can manifest at both radio and optical wavelengths as complex, asymmetric distribution of the brightness over the source map,
spatial bias between the optical and radio brightness centroids, core-shift effects, spatial bias between the core/AGN brightness centroid and
with respect to the optical centroid of the host galaxy.
Moreover, the structure effects are often variable.
Although many studies are devoted to this factor, see,  
\citet{Fey1997,SilvaNeto2002,Moor2011,Bouffet2013,Zacharias2014,Berghea2016} and papers cited therein,
there are insufficient data to quantify the impact of source structure on the  accuracy of the orientation angles between optical and radio frames.
Evidently, the most complete study is provided by \citet{Zacharias2014} used in the current work. 
It should be noted that though the structure effect can reach several mas for an individual source, it can hardly be systematic and
thus will be averaged over the sky during the computation of the orientation angles between the GCRF and ICRF.
However, supplementary observations and theoretical considerations are needed to quantify this effect more accurately.  

Two main methods to obtain the link between the GCRF and ICRF were considered in this paper.
The first method is direct {\it Gaia} observations of the sufficiently optically bright ICRF sources.
This method allows for a straightforward solution of the task. 
However, there are serious constraints on the accuracy of this method caused by the previously mentioned factors.
These factors can limit the real accuracy of the GCRF--ICRF link to 0.1 mas or worse.

Observations of radio stars can serve as an alternative equally accurate method.
It was successfully used to link the HCRF to the ICRF.
However this method is also affected by some accuracy-limiting factors \citep{Malkin2016e}.
Many radio stars comprise double or multiple systems, and thus their orbital motions must be accounted for.
The accurate {\it Gaia}-derived orbits can be used for this purpose.
Moreover, radio stars may have complex and variable structures, which might cause a time-dependent bias between the radio and optical positions.
\citet{Lestrade1995} estimated the impact of radio star structure and possible variations in the radio star emitting centre to be within
the error budget of approximately 0.5 mas.
\citet{Lestrade1999} found that the structure-induced systematic errors in the VLBI positions of 12 stars ranged from 0.07~mas to~0.54 mas,
with a median value of 0.18~mas.
Provided that several tens of radio stars have been observed, this factor should not significantly impact on the errors in the orientation angles. 

Combination of two methods of linking the GCRF to the ICRF should facilitate improvement of the systematic accuracy of the link between radio and optical frames.

Finally, it should be noted that the preparation for the aligning of the {\it Gaia} catalog to ICRF3 planned for 2018--2019 is only an intermediate stage
in construction of the multi-frequency celestial reference frame.
Certainly, the main work on the link between radio and optical frames is to be done in the early 2020s, after the final {\it Gaia} catalog
is prepared.
It is desirable to plan preparation of a new ICRF realization, will it be called ICRF4 or ICRF3 extension, at the same time,
i.e. immediately before the final GCRF link to ICRF.
Comparison of gsf2015b with ICRF2 has clearly shown that the latest radio catalog should be used for this work due to the accuracy of 
the VLBI-derived CRF solutions which rapidly improves with time.
In the framework of this activity, it appears to be very important to fast-track the plans on the ICRF improvements in the Southern Hemisphere
\citep{Jacobs2014,Plank2015}, and to start a program of observations of radio stars \citep{Malkin2016e}.

As the {\it Hipparcos} experience has shown, it can be expected that the work on improving the link between radio and optical frames
will be continued during a long period after completing the {\it Gaia} mission.
Continuous quality improvement of a VLBI-based ICRF promises corresponding improvement of this link with time.


\section*{Acknowledgements}

The author would like to thank the two reviewers and the handling editor for their valuable help in improving the manuscript.
This work was partially supported by the Russian Government Program of Competitive Growth of Kazan Federal University.
This research has made use of NASA's Astrophysics Data System.


\bibliography{zm_icrf-gcrf_v3}
\bibliographystyle{frontiersinSCNS_ENG_HUMS}

\end{document}